\documentclass[fleqn,twoside]{article}
\usepackage{espcrc2}
\usepackage{epsfig}

\usepackage{graphicx}
\usepackage[figuresright]{rotating}

\newcommand{\AmS}{{\protect\the\textfont2
  A\kern-.1667em\lower.5ex\hbox{M}\kern-.125emS}}
\newcommand{\lsim}{\stackrel{<}{_\sim}}
\newcommand{\ra}{\rightarrow}
\newcommand{\be}{\begin{equation}}
\newcommand{\ee}{\end{equation}}
\newcommand{\bea}{\begin{eqnarray}}
\newcommand{\eea}{\end{eqnarray}}
\newcommand{\non}{\nonumber}

\newcommand{\ket}[1]{\,\left|\,{#1}\right\rangle}

\title{Intrinsic Charm in B-Meson Decays
{\small \hfill{SLAC-PUB-9490} \\ 
 \rightline{UK/TP 02-14}}
}
\author{S. Gardner
\address{
Stanford Linear Accelerator Center, Stanford University, Stanford, 
CA 94309 USA }
\address{
Department of Physics \& Astronomy, University of Kentucky, Lexington, 
KY 40506-0055 USA }
\thanks{Invited talk at the 5th International
Conference on Hyperons, Charm and Beauty Hadrons (BEACH 2002),
University of British Columbia, Vancouver, June 25-29, 2002.}
}

\begin{document}

\begin{abstract}
Light cone hadron wave functions support Fock 
states of arbitrarily high particle number: 
their heavy quark content arises naturally through QCD interactions.
We discuss what role $c\bar c$ pairs, intrinsic to 
a hadron's structure, can play in B-meson decays. 
The effects can be prominent in hadronic decays for which the tree-level
contributions are Cabibbo-suppressed, as in $B\to \pi K$ 
decay, and they mimic ``charming penguin'' contributions. 
\vspace{1pc}
\end{abstract}

\maketitle

\section{INTRODUCTION}

It is usually assumed in the analysis of
$B$-meson decays that only the valence quarks of the initial and
final-state hadrons participate in the weak transition. 
Any non-valence gluon or sea quarks present in the
initial or final state wave functions appear only as spectators.
However, the wave functions of a bound state in a relativistic quantum field
theory such as QCD necessarily contain Fock states of arbitrarily high
particle number. This takes on new significance in light of 
the hierarchical structure of the CKM matrix
--- the weak transition $b\ra s c{\overline c}$ is doubly Cabibbo
enhanced with respect to a $b\ra s u{\overline u}$ transition.
The small probability of realizing a Fock state containing a $c \bar c$ pair 
is offset by the comparatively large CKM matrix
elements associated with the $b\to s c {\overline c}$ transition,
promoting its phenomenological impact. In this talk, I discuss 
how the presence of intrinsic charm in the hadrons' light-cone wave 
functions, even at a few percent level, can impact B-meson decays.  
My remarks are based on work done in collaboration 
with Stan Brodsky~\cite{sjbsvg}. 

To begin, we define intrinsic charm (IC) and  
review the evidence for its presence 
in the light hadrons, arguing that the magnitude of IC 
could be larger in the B-meson. 
We proceed to consider the disparate roles for IC in B physics, 
reviewing how it can mediate certain, rare decays, how it can enter 
the semi-leptonic branching ratio cum ``charm counting'' puzzle, 
and finally how it can act as a source of hadronic pollution, 
impacting the extraction of $\gamma$ in $B\to \pi K$ decays. 

\section{INTRINSIC CHARM?}

\begin{figure}
\includegraphics[scale=0.43]{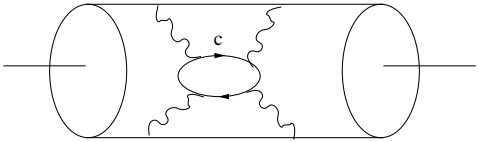} 

\vspace{-9mm}
\caption{An intrinsic charm contribution to a meson's self-energy.}
\label{fig:intrin}
\end{figure}

The wave functions of a relativistic bound state
contain Fock states of arbitrarily high particle number. For example, 
\bea
&\!\!\! \ket{B^- } =
\psi_{b \bar u} \ket{ b \bar u }+
\psi_{b \bar u g} \ket{ b \bar u g } +
\psi_{b \bar u d \bar d} \ket{ b \bar u d \bar d } \non \\
&\!\!\! + \psi_{b \bar u c \bar c} \ket{ b \bar u c \bar c } + \cdots.
\eea
The Fock state decomposition is usually performed at equal light-cone
time using light-cone quantization in 
light-cone gauge $A^+ =0$~\cite{Dirac:1949cp,Brodsky:1989pv}.
The non-valence partons of the 
higher Fock states arise from QCD interactions. The partons of a
Fock component are entangled
through multiple gluon interactions, as illustrated in Fig.~\ref{fig:intrin}; 
this makes them 
{\it intrinsic} to the hadron's structure. 
The intrinsic, heavy quarks are part of the
hadron's non-perturbative bound-state~\cite{Brodsky:1980pb}.

\begin{figure}

\vspace*{-7mm}
\includegraphics[scale=0.43]{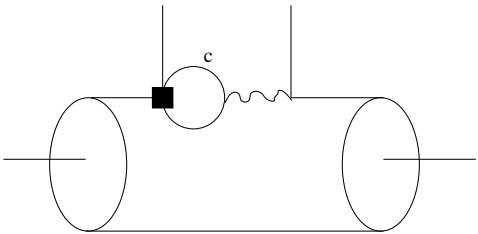} 

\vspace{-9mm}
\caption{Extrinsic charm as a radiative correction to the 
effective Hamiltonian, whose action is denoted by the
square box, for $B\to h_1 h_2$ decay.}
\label{fig:peng}
\end{figure}
In contrast, a perturbative correction to the weak transition 
matrix element can yield a $c\bar c$ pair through gluon splitting; the
quark pair is generally not multiply connected to the partons of the 
bound state and thus is {\it extrinsic} to the hadron's structure, 
as illustrated in Fig.~\ref{fig:peng}. 

Generally, 
``intrinsic'' contributions in B-meson decay
are of higher twist than the leading contribution, 
whereas ``extrinsic'' contributions
are of higher order in $\alpha_s$: they may not be 
crisply separable. 

\subsection{Evidence for intrinsic charm}

Deeply inelastic scattering (DIS) measurements at large momentum 
transfer, $Q^2 \gg 1 \;(\hbox{GeV/c})^2$, reveal the proton 
to have a rich sea structure. 
Our interest is in observables which can distinguish the 
perturbative evolution of the proton's structure functions from 
``intrinsic'' effects. For example, the proton's flavor structure permits
it to fluctuate to $K^+\Lambda$, making 
the $s$ and $\bar s$ parton distribution functions unequal~\cite{strange}.  
This effect has recently been observed~\cite{nutev}, so that 
intrinsic effects are appreciable, 
but what of intrinsic charm? 
The intrinsic heavy-quark 
fluctuations in hadrons can be analyzed using the operator-product
expansion: Franz et al. determine that 
the momentum fraction carried by
the heavy $Q\bar Q$ pair scales as $1 / m^2_Q$~\cite{Franz:2000ee},
where we assume that the associated hadronic matrix element is characterized
by $k_\perp^2$, where 
$k_\perp$ is a typical momentum scale in the hadron light-cone 
wave function. In
contrast, in Abelian theories,
the contribution of an intrinsic, heavy lepton pair
to the bound state's structure first appears in
${\cal O}(1/m_L^4)$. Franz et al. estimate that the intrinsic charm
probability in the proton is $\lsim 1\%$~\cite{Franz:2000ee}. 
Analyses of DIS data are consistent
with this prediction~\cite{disic}. 
Phenomenologically, the presence of IC at the $\lsim 1\%$ level 
is appealing. IC naturally
explains the ``$\rho\pi$ puzzle'' in $J/\psi (\psi^\prime)$ 
decays~\cite{puzzle},
and an explicit IC component is needed to describe the $x_F\to 1$
charm production and polarization data in $\pi N$ reactions~\cite{leading}. 

\subsection{Intrinsic charm in the B-meson}

The existence of intrinsic charm (IC) in the proton implies that
IC exists in other hadrons as well. 
In order to translate an estimate of the IC probability in the
proton  to that of the IC of a $B$-meson, we
are faced with two conflicting effects. The typical internal
transverse momentum $k_\perp$
is larger in the B-meson, evidently favoring a larger IC
probability in the B meson; on the other hand, 
the proton's additional valence quark generates
a larger combinatoric number of IC diagrams,
favoring a larger IC probability
in the proton. The magnitude of $k_\perp$ is significantly
larger in the B-meson; $\lambda_1$, the kinetic energy of the b quark
in the B-meson, is 
$\sqrt{|\lambda_1|}\sim 0.4$ GeV~\cite{lambda1}. 
Minimizing the light-cone energy
of the partons in the IC Fock component of the free light-cone
Hamiltonian lead Chang and Hou to estimate that the momentum
fraction carried by charm is smaller in the B-meson, namely
$\langle x_c \rangle \approx 0.22$, than in the proton, for which 
$\langle x_c \rangle \approx 0.28$~\cite{Chang:2001iy}. 
Thus the IC probability must be
higher in the B-meson to make the charmed quarks carry the same, fixed fraction
of the hadron's momentum. 
These estimates suggest that the IC probability in the B-meson 
could be as large as a few percent, and the presence of IC 
in the $\Lambda_b$
baryon could be larger still. 

\section{ROLES FOR INTRINSIC CHARM IN B-DECAY}

Intrinsic charm can play a variety of roles in B-meson decays: it
can act as a mediator of certain, rare decays; it may give new insight
on old (resolved?) puzzles;
and it may impact the extraction of CKM information 
from decays to strange, charmless final states. We consider each
of these roles in turn. 

The presence of intrinsic charm quarks in the $B$ wave function provides
new mechanisms for $B$ decays.  For example, 
the production of final states with three charmed quarks, 
$b\to c\bar c c X$, such as $\bar B \to 
J/\psi D \pi$ and $\bar B \to J/\psi D^{(*)}$~\cite{Chang:2001iy}, 
are difficult to realize in a valence model, as the 
$c\bar c$ pair can only be realized through OZI-violating processes. 
They occur naturally, however, when the
$b$ quark of the intrinsic charm Fock state $\ket{ b \bar u c \bar c}$
decays via $b \to c \bar u d$ --- the intrinsic $c\bar c$ component of the
B-meson is materialized in the final state. Chang and Hou suggest that
the slight excess in the inclusive $B\to J/\psi X$ yield at low $J/\psi$
momentum, observed by CLEO~\cite{CLEO}, hints to the presence 
of $\bar B \to J/\psi D^{(*)}$~\cite{Chang:2001iy},
though such an effect could also be generated by
$\bar B\to J/\psi \Lambda \bar n$ decay~\cite{Brodsky:1997yr}. 
Intrinsic charm in the B meson can also mediate $B\to J/\psi \gamma$
and $B\to J/\psi e^- \bar \nu_e$ decays. Numerical estimates have
been made for $\bar B^0 \to J/\psi X e^- \bar \nu_e$ and 
$\bar B^0 \to D \bar D X e^- \bar \nu_e$ decay; the branching ratios 
are markedly larger when IC in the B-meson is included, though they
remain small, note, e.g., 
${\cal B}(B^- \to J/\psi e^- \bar \nu_e X) 
\approx 4 \cdot 10^{-7}$~\cite{isuic}. 

Intrinsic charm could well prove helpful in resolving long-standing
puzzles in B-physics. For example, the observed lifetime 
ratio of the $B$ and $\Lambda_b$ hadrons, 
$\tau(\Lambda_b)/\tau(B)|_{\rm expt}=0.797 \pm 0.053$~\cite{pdg02},
differs significantly from unity, the result predicted
in the heavy-quark limit. Spectator effects, i.e.,
the manner in which the decaying $b$ quark interacts with its
hadronic environment, 
are evidently crucial to explaining the discrepancy~\cite{spectator}; 
spectator interactions involving intrinsic charm ought differ 
in the $B$ and $\Lambda_b$ and could play a role.

\begin{figure}
\includegraphics[scale=0.8]{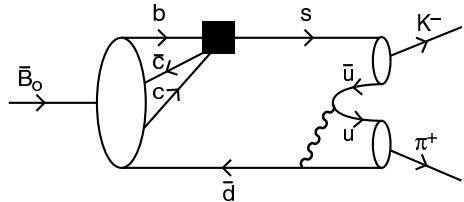} 

\vspace*{-10mm}
\caption{Intrinsic charm in the B-meson can mediate the decay to a
strange, charmless final state via the weak transition
$b\to s c {\overline c}$.}
\label{fig:icdiag}
\end{figure}

Moreover, the semileptonic branching
fraction in inclusive B-meson decay, B$_{\rm sl}$, is a bit low with
respect to SM predictions; however, the 
``natural'' resolution of this puzzle --- an increased
$b\to s c{\bar c}$ rate ---  is untenable, as the 
observed number of charm (and anti-charm) quarks per B-meson decay, 
$n_c$, is consistent with SM expectations~\cite{waller}. 
That is, B$_{\rm sl}$ would decrease were the hadronic width of
the B-meson to increase, but the hadronic width is tied to $n_c$, the
average yield of $c$ and $\bar c$ per B-meson decay. 
IC in the B-meson can increase the charmless decay
rate, as in the exclusive process shown in Fig.~\ref{fig:icdiag},
thus reducing the semileptonic branching ratio. IC thus acts to 
loosen the correlation between the hadronic width, particularly
the $b\to s c \bar c$ rate, and $n_c$. 
Earlier work ascribed a possible role to IC in resolving
the $B_{\rm sl}$ cum $n_c$ puzzle~\cite{Dunietz:1998zn}, yet only
IC in the light hadrons was considered. 
The role played by IC in the B meson in realizing strange,
charmless final states may be of greater importance.

Finally, IC could be important to understanding the 
empirical $B\to \eta^\prime K\;,\eta^\prime X$ branching ratios, 
which are large with respect to SM estimates. 
Previously, a 
valence $c\bar c$ component in the $\eta^\prime$ had
been invoked to resolve the disparity~\cite{Halperin:1997as}, but
the decay constant 
$f_{\eta^\prime}^{(c)}$, namely 
$\langle 0 | {\bar c}\gamma_\mu \gamma_5 c| \eta^\prime(p)\rangle
\equiv i f_{\eta^\prime}^{(c)} p_\mu$, 
is too small~\cite{Franz:2000ee}; 
efforts to reconcile the observed
rate with SM predictions continue. 
Although other mechanisms could well
be at work~\cite{np}, the factorization approximation
does not capture the physics of IC.
IC is produced in a higher Fock component of a hadron's light-cone
wave function; it is naturally in a color octet state~\cite{yuan}, 
so that the dynamical
role it plays in mediating B-meson decay is intrinsically
non-factorizable in nature.

\subsection{Intrinsic charm in $B\to \pi K$ decays}

We now turn to the role of intrinsic charm in $B\to\pi K$ decay. 
These decays are penguin dominated as $b\to s u \bar u$ decay 
is ${\cal O}(\lambda^2)$ suppressed. 
The $|\Delta B|=1$ effective Hamiltonian for $b\to s q{\bar q}$ decay is
\bea
&{\cal H}_{\rm eff} = \frac{G_F}{\sqrt{2}} \{
\sum_{p=u,c} [ V_{pb}V_{ps}^\ast(C_1{O}_1^{p} + C_2{O}_2^{p})] \non \\
& - V_{tb}V_{ts}^\ast \sum_{j} C_j {O}_j \} \;,
\label{heff}
\eea
where $C_2 \sim {\cal O}(1)$. Sums of the 
products of $C_i(\mu)$ and the matrix elements of $O_i(\mu)$ are
renormalization scale and scheme invariant, so that 
each such entity is a parameter: the set constitutes 
a parametrization~\cite{Buras:2000ra}.
For $B\to \pi K$ decay, in valence approximation, 
the decay amplitude ${\cal A}(B^0 \to K^+\pi^-)$, e.g., 
can be written~\cite{Buras:2000ra}
\bea
&{\cal A}(B^0 \to K^+\pi^-) =
V_{us}V_{ub}^\ast[
E_1  - P_1^{\rm GIM}] \non\\
&- V_{ts}V_{tb}^\ast P_1 \;.
\label{burasparam}
\eea
The parameter $E_1$ contains 
$W^+$ emission topologies, 
$P_1$ contains penguin topologies, and 
$P_1^{\rm GIM}$ contains penguin contributions which
vanish in the $m_c=m_u$ limit.
Beyond valence approximation, 
the form of the parametrization does not change,
but additional contributions arise. That is, from IC, we have
the additive contribution $V_{cs} V_{cb}^\ast A_1^{\rm IC}$, where
\be
A_1^{\rm IC} = C_1 \langle O_1^c \rangle^{\rm IC}
+ C_2 \langle O_2^c \rangle^{\rm IC} \;,
\ee
to ${\cal A}(B^0 \to K^+ \pi^-)$ --- 
a contribution of this ilk is illustrated in Fig. \ref{fig:icdiag}. 
The IC contribution is Cabibbo-enhanced and 
contains an ${\cal O}(1)$ Wilson coefficient, just as 
``charming penguins'' do~\cite{charming,isola}, so that charming penguins
need not be penguins at all. 
Indeed, the parameter 
$A_1^{\rm IC}$ can be absorbed into $P_1$ and $P_1^{\rm GIM}$. 
If the contributions are driven by non-perturbative physics, 
as Ciuchini et al. argue~\cite{charming}, 
the contributions are probably indistinguishable. We now proceed
to the evaluation of the parameters we have introduced.

In the usual perturbative QCD treatment of exclusive processes,
the amplitude for a particular exclusive process is formed
by the convolution of the nonperturbative distribution amplitudes,
$\phi_H(x,Q)$, with the hard scattering amplitude, $T_H$, computed
from the scattering of on-shell, collinear
quarks~\cite{Lepage:1979zb,Lepage:1980fj}. For $B\to M_1 M_2$ decay,
we have
\bea
\label{classic}
& {\cal M}(B \to M_1 M_2) = \int_0^1 dz \int_0^1 dy \int_0^1 dx
\phi_B(x,Q) \non \\
& T_H(x,y,z) \phi_{M_1}(y,Q) \phi_{M_2}(z,Q) \;,
\eea
where, e.g.,
$\phi_{M_2}(z,Q)=\int_0^{Q} d^2 k_\perp \phi(x,k_\perp,\lambda_i)$.
This formula is suitable if the
distribution amplitudes vanish
sufficiently rapidly at the endpoints, and if $\alpha_s(\mu)$ is sufficiently
small for a perturbative treatment to be germane; however, these 
criteria do not appear to be satisfied in 
$B\to \pi l \nu_l$ decay~\cite{Szczepaniak:1990dt,Burdman:1991hg}. 
Equation (\ref{classic}) itself emerges from an expansion
of $T_H$ in powers of $k_\perp^2/Q^2$; one solution~\cite{Botts:1989kf}
is to reorganize the contributions in $k_\perp$, so that the
contributions to the hard scattering in the transverse configuration
space ($b$, conjugate to $k_\perp$) are no longer of point-like size.
The $b$ dependence of the reorganized distribution amplitudes,
the so-called ``Sudakov exponent,'' suppresses the large $b$ region,
so that the resulting integrals are convergent and $\alpha_s(\mu)$
is more or less consistently small. These ideas have been extended
to B-meson decays by Li and collaborators: 
the $B\to \pi$ form factor is regarded as perturbatively calculable,
once proper resummation techniques are applied~\cite{hnli}. 

In the approach of Beneke {\it et al.}~\cite{Beneke:2001ev}, however,
the $B\to \pi$ form factor is treated as
non-perturbative input. One consequence is that
the perturbative corrections in the two schemes are
organized differently:
in Ref.~\cite{Beneke:2001ev}, hard-scattering contributions in
${\cal O}(\alpha_s)$ are retained
which would be regarded as non-leading order in the approach
of Ref.~\cite{hnli}. However, infrared enhancements also plague
the treatment of annihilation contributions in this
approach~\cite{Beneke:2001ev};
thus the treatment of Ref.~\cite{hnli} is more systematic
in that the contributions of all the decay
topologies are regulated in the same way. 
The internal consistency of 
the Li et al. framework has, however, been criticized~\cite{sjbsvg,dgs}. 
We shall nevertheless adopt it in the discussion to follow. 

\begin{figure}
\includegraphics[scale=0.28]{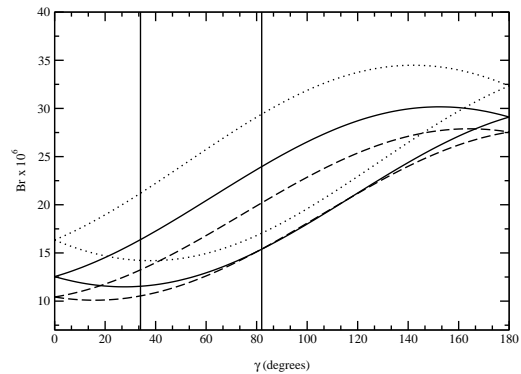}

\vspace*{-10mm}
\caption{
The impact of IC in the B meson on the
$B\to K^\pm\pi^\mp$ branching ratios 
as a function of $\gamma$. The {\it upper} curve for each type
of line corresponds to $B^0\to K^+\pi^-$ decay, whereas
the {\it lower} curve for each type of line corresponds to
${\bar B}^0\to K^- \pi^+$ decay. The solid line depicts the
results of Ref.~\protect\cite{lks}.
The dashed line is the result once the IC contribution,
is subtracted, and the dotted line
is the result once the IC contribution is added.
The vertical solid lines enclose the $\ge 5\%$ C.L.
fits to $\gamma$ in the SM,
$34^\circ \ge \gamma \ge 82^\circ$, of Ref.~\protect\cite{Hocker:2001xe}.}
\label{fig:keum}
\end{figure}

Adopting the notation and conventions of Ref.~\cite{lks}, 
the parameters of Eq.~(\ref{burasparam}) 
can be mapped to 
\begin{eqnarray}
& E_1 = -f_K F_e - M_e \quad ; \quad P_1^{\rm GIM} = 0 \,, 
\nonumber \\
& P_1 =  -f_K F_e^P - M_e^P - f_B F_{a}^P - M_{a}^P \,,
\end{eqnarray}
where ``factorizable'' and ``non-factorizable'' contributions
are denoted by $F$ and $M$, respectively. The subscripts $e$ and $a$ 
refers to emission and annihilation topologies, respectively, and the
$P$ superscript reflects the presence of penguin operators 
in the hard-scattering amplitude. The $F$ and $M$ form factors
are calculated in leading order in $\alpha_s$; in this order,
the ``extrinsic charm'' 
contribution of Fig. \ref{fig:peng} does not occur. The results of including
the IC contribution, as per Ref.~\cite{sjbsvg}, are 
shown in Fig.~\ref{fig:keum}. One striking feature 
of this approach is the large direct CP asymmetry 
predicted in $B^0 \to K^+ \pi^-$ decay~\cite{lks,Keum:update}, 
in distinction to the predictions of Ref.~\cite{Beneke:2001ev}. 
Experiment does not currently favor either scenario~\cite{directcp}. 
Note that IC can act to either enhance or 
decrease the CP asymmetry; 
IC, or as yet uncomputed NLO effects, could mute the
distinction between the two approaches. Moreover, the 
IC contribution we estimate, $|A_1^{\rm IC}|/|P_1| \sim {\cal O}(10)\%$, 
is a non-trivial fraction of the penguin parameter
$P_1$; its presence casts doubt on the ability to calculate
the effective value of $P_1$. 
Moreover, a factorization assumption for the
effective penguin contribution in $B^+ \to K_0 \pi^+$ decay, needed for 
an SU(3)$_f$-based analysis of the time-dependent asymmetries
in $B\to \pi^+\pi^-$ for $\sin(2\alpha)$~\cite{groros}, may
not be warranted.

\section{SUMMARY}

The role of non-valence components in the hadrons' light-cone
wavefunctions, coupled with the hierarchical structure of
the CKM matrix, offers new perspective  on B decays. 
The effects can be striking in exclusive 
decays with Cabibbo-suppressed tree contributions, thus our focus on 
intrinsic charm in $B\to \pi K$ decay. Intrinsic charm effects
are quantitatively elusive, but can be expected to play a non-trivial
role in decays to strange, charmless final states. They may prove
inseparable from 
charming penguin contributions. Such effects can modify the
effective value of $P_1$, the penguin contribution, from its expected  value, 
to confound the determination of
$\gamma$. The observation of the rare decay $B\to J/\psi D X$ decay, or
any of its brethren, would be clarifying.

\vspace{4mm}
\noindent
{\bf ACKNOWLEDGEMENTS}
It is a pleasure to acknowledge Stan Brodsky for an engaging and
fruitful collaboration, as well as the SLAC Theory Group for
its hospitality. I thank the conference organizers for the 
invitation to speak and 
the U.S. Department of Energy for support under contracts 
DE-FG02-96ER40989 and DE-AC03-76SF00515.

\end{document}